\documentclass{article}

\usepackage[utf8]{inputenc} 
\usepackage[T1]{fontenc}    
\usepackage{hyperref}       
\usepackage{url}            
\usepackage{booktabs}       
\usepackage{amsfonts}       
\usepackage{nicefrac}       
\usepackage{microtype}      
\usepackage{lipsum}
\usepackage{graphicx}
\usepackage[tbtags]{amsmath}
\usepackage{amsthm}
\usepackage{amsfonts}
\usepackage{amssymb}
\usepackage{latexsym,bm}
\usepackage[english]{babel}
\usepackage{fancyhdr}
\usepackage{multirow}
\usepackage{indentfirst}
\usepackage{graphicx}
\usepackage{color}
\usepackage{balance}
\usepackage{authblk}
\usepackage{titlesec}
\usepackage{booktabs}
\usepackage{caption2}
\usepackage[T1]{fontenc}

\usepackage{array}
\usepackage{algorithm}
\usepackage{algorithmic}
\newtheorem{definition}{Definition}

\title{Applying Rational Envelope curves for skinning purposes}

\author{
	Kinga Kruppa\thanks{} \\
	University of Debrecen, Doctoral School of Informatics\\
	University of Debrecen, Faculty of Informatics, Debrecen, H-4028, Hungary\\
	\texttt{kruppa.kinga@inf.unideb.hu} \\
}

\begin{document}
\date{}
	\maketitle
	\begin{abstract}
		Special curves in the Minkowski space such as Minkowski Pythagorean hodographs play an important role in Computer Aided Geometric Design, and their usages have been thoroughly studied in the recent years. Also, several papers have been published which describe methods for interpolating Hermite data in $\mathbb{R}^{2,1}$ by MPH curves. \cite{Bizzarri2016} introduced the class of RE curves and presented an interpolation method for $G^{1}$ Hermite data, where the resulting RE curve yields a rational boundary for the represented domain. We now propose a new application area for RE curves: skinning of a discrete set of input circles. We find the appropriate Hermite data to interpolate so that the obtained rational envelope curves touch each circle at previously defined points of contact. This way we overcome the problematic scenarios when the location of the touching points would not be appropriate for skinning purposes.
	\end{abstract}

	
	\section{Introduction}\label{section:introduction}
	
	Medial Axis Transform (MAT) is a widely studied area both in computer graphics and in image processing. For a given planar domain, the Medial Axis (MA) is the locus of the centers of the maximal inscribed disks, and the MAT can be obtained by lifting the MA into the space by the radii of the inscribed disks. Once we obtain the MAT of a domain, we can reconstruct the boundary of the domain by using the well-known envelope formula by \cite{Choi1997, Choi1999}. Curves in the Minkowski space are suitable to describe MATs, however, only special curves describe domains whose boundaries are rational. It was shown by \cite{Moon1999} that Minkowski Pythagorean curves are such, and the yielded envelope curves are not only rational but Pythagorean hodograph curves, thus, their offsets are rational as well. 
	
	Nowadays, several Hermite interpolation methods have been published using MPH curves so that the resulting envelope is rational (\cite{Kim2003,Kosinka2006,Kosinka2009,Kosinka2010,Kosinka2011,Bizzarri2019}). In 2016, \cite{Bizzarri2016} showed that a broader class of curves exists in $\mathbb{R}^{2,1}$ that yield rational boundaries: the so-called Rational Envelope (RE) curves. Thus, if only the rationality of the envelope is required, one can rely on RE curves (which are simpler to compute), while one has to restrict oneself to MPH curves only if the rationality of the offsets are needed as well. The authors described a $G^1$ interpolation method to construct an RE interpolant, which can be used to create rational blending surfaces between canal surfaces.

	Now let us introduce a quite different area in Computer Aided Geometric Design: skinning. Standard point-based curve and surface modeling has always played an important role in CAGD, but there has been a growing interest in curve and surface modeling based on different objects, such as circles and spheres. Skinning has become known as constructing a pair of at least $G^1$ continuous splines for a predefined sequence of circles which touch each circle at one point. The idea can also be extended to 3D, leading to the skinning of an input set of spheres. In recent years, several skinning methods have been published, such as \cite{Kunkli2010, Bana2014, Bastl2015, Kruppa2019}. 
	
	From the outlook, skinning and boundary reconstruction are somehow similar, but it is important to note that the problem settings are fundamentally different. In skinning, we cannot talk about the reconstruction of the boundary, since there is no initial domain assumed at all. Despite this fact, based on the similarities of the two es, we propose to approach skinning from an aspect of MAT and use the yielded rational envelopes.  
	
	In this paper, we introduce how we apply the MPH/RE interpolation method proposed by \cite{Bizzarri2016} for the aim of skinning, and we show a geometric construction for appropriately choosing the Hermite input data. We approach it in a somehow inverse way: instead of having the tangent vectors as input, we fix the location of the touching points, from which we then calculate the tangent vectors. In Section \ref{section:previouswork},  we overview the two methods that stand as a basis for our work. Then, in Section \ref{section:oursolution} we introduce our novel approach to use RE curves as skinning, and we show some examples as well. Finally, Section \ref{section:conclusion} concludes our results.

	\section{Motivation and related work}\label{section:previouswork}
	
	Firstly, let us consider the  parametric curve $\overline{\textbf{y}}(t) = (\textbf{y}(t), r(t))$ as the MAT of a domain. Then the boundary (envelope) of the domain can be reconstructed using the envelope formula (\cite{Choi1997, Choi1999}):
	\begin{equation}\label{eq:envelope}
	\begin{split}
	&\textbf{x}^{\pm}( t)  =
	\textbf{y} ( t ) -\\
	& r (t ) \frac{ r'(t)\ \textbf{y}'(t)  \pm \textbf{ y}'^\perp (t)\sqrt{||\textbf{y}'(t)||^2 - r'(t)^2 }}{||\textbf{y}'(t) ||^2}
	\end{split}
	\end{equation}
	
	Now let us overview the $G^1$ Hermite interpolation method proposed by \cite{Bizzarri2016}, since we will use this RE curve construction method for skinning approach. In the second part of this section, we examine the standard skinning method of \cite{Kunkli2010}.
	
	\subsection{$G^1$ Hermite interpolation yielding rational envelopes}\label{subsection:mat}
	
	Given two points and two vectors as the input Hermite data to interpolate, let $P_0$ and $P_{1}$ note the end points, and let $\textbf{t}_0$ and $\textbf{t}_{1}$ note the end tangent vectors. Let us mark projection as $\overset{\triangledown}{\textbf{n}} = (n_x, n_y)$ if $\textbf{n}=(n_x, n_y, n_z)$ and $\textbf{n} \in \mathbb{R}^{2,1}$. Let us mark rotation as $\textbf{n}^{\perp}= (n_y, -n_x)$ if $\textbf{n}=(n_x, n_y)$ and $\textbf{n} \in \mathbb{R}^{2}$.
	
	By directly applying the envelope formula \ref{eq:envelope}, we can get the touching points $Q^\pm_i$ ($i \in \{0,1\})$ of the corresponding envelope curve as:
	\begin{equation}\label{eq:env_q}
	Q_i^{\pm} = \overset{\triangledown}{P_i} - P_{i_z} \frac{t_{i_z} \overset{\triangledown}{\textbf{t}_i} \pm \overset{\triangledown}{\textbf{t}_i}^\perp \sqrt{||\overset{\triangledown}{\textbf{t}_i}||^2 - t_{i_z}^2}}{||\overset{\triangledown}{\textbf{t}_i}||^2}
	\end{equation}
	
	The construction relies only on one branch, and we define the touching point $Q_i$ as $Q^+_i$. We then define appropriate tangent vectors:
	\begin{equation}\label{eq:env_v}
	\textbf{v}_i = \alpha_i \frac{{(\overset{\triangledown}{P_i} - Q_i)}^\perp}{||{\overset{\triangledown}{P_i} - Q_i}||} \qquad  \alpha_i \in \mathbb{R}
	\end{equation}

	Once we obtain $Q_i$ and $\textbf{v}_i$, we construct the planar curve $\textbf{x}(t)$, $t \in [0,1]$ that interpolates $Q_i$ and $\textbf{v}_i$, for example with Hermite interpolation.
	
\begin{figure}
		\centering
		\includegraphics[width=0.7\linewidth,, trim={1.5cm 4cm 1.9cm 2.4cm},clip]{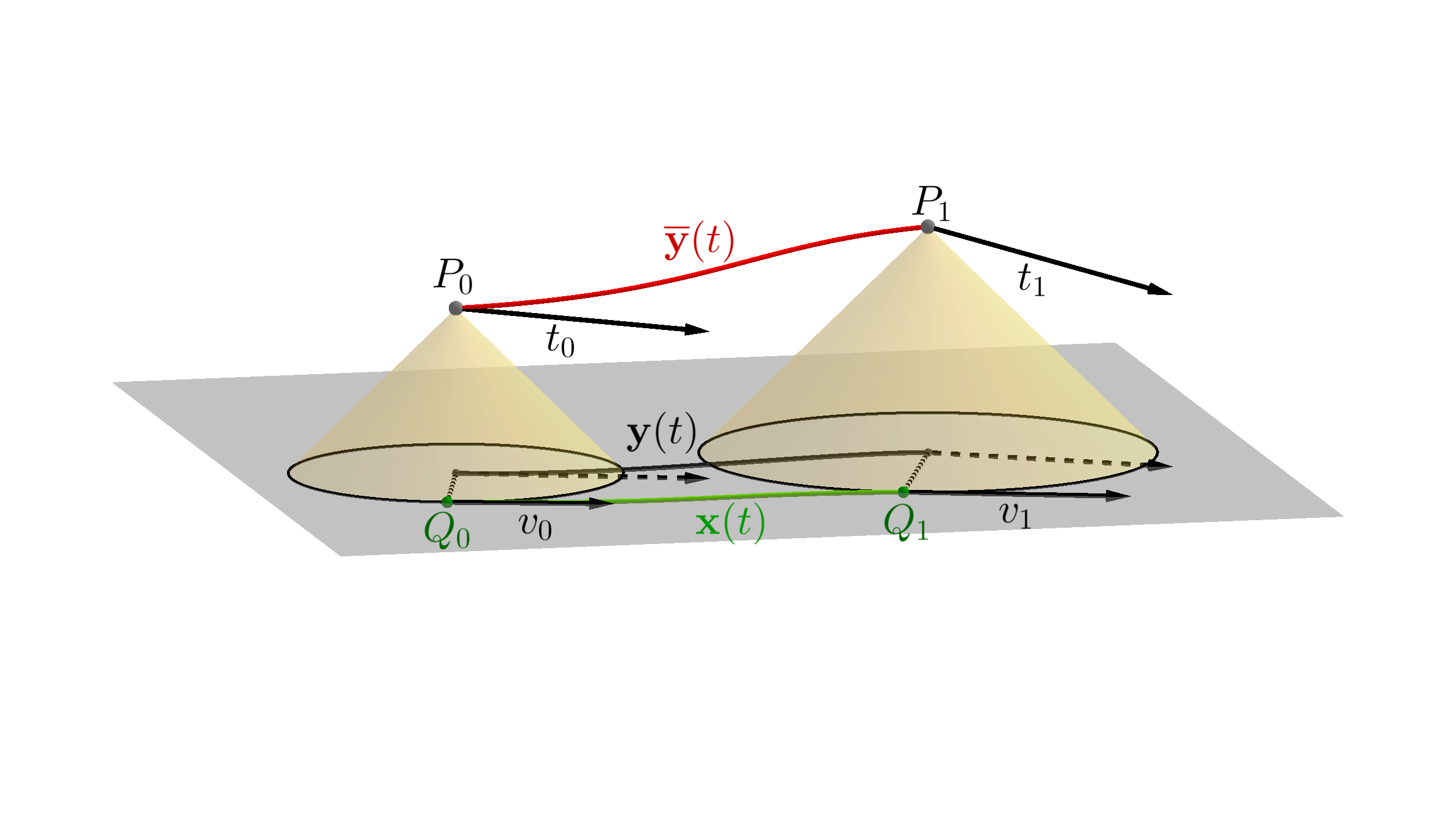}
		\caption{Construction of the RE curve $\overline{\textbf{y}}(t)$ which interpolates the input data $P_i$ and $t_i$ ($i \in \{0,1\}$) with  $G^1$ continuity, using the method of \cite{Bizzarri2016}.}
		\label{fig:bizzarri-ybar}
	\end{figure}

	The final step is to construct the medial axis $\textbf{y}(t)$ as a one-sided offset of  $\textbf{x}(t)$ with varying distance:
	$$ \textbf{y}(t) = \textbf{x}(t) + r(t) \frac{\textbf{x}'^\perp(t)}{||\textbf{x}'(t)||} 
	=  \textbf{x}(t) + f(t)\  \textbf{x}'^\perp(t)$$
	which is rational only if
	$ f(t) = \frac{r(t)}{||\textbf{x}'(t)||} $ 
	is a rational function. To assure that the resulting curve interpolates the initial input data $P_i$ and $\textbf{t}_i$, $f(t): [0,1] \rightarrow \mathbb{R}$ has to be constructed as a polynomial function with the following boundary conditions ($i \in \{0,1\}$):
	\begin{equation}\label{eq:fcondition}
	\begin{split}
	&f(i) = \frac{P_{i_z}}{||\textbf{v}_i||} \\
	&f'(i) = - \frac{\overset{\triangledown}{t_i} \cdot (f(i)\ \textbf{x}''(i) - v_i^\perp)}{\overset{\triangledown}{t_i} \cdot \textbf{v}_i} 
	\end{split}
	\end{equation}
	By lifting the medial axis back to the space, the final form of the interpolant is:
	\begin{equation}\label{eq:re_curve}
	\begin{split}
	\overline{\textbf{y}}(t) = &\left(\textbf{x}_x(t) + \textbf{x}'_{y}(t) \ f(t) ,\  \textbf{x}_y(t) -\textbf{x}'_x(t) \ f(t) ,\  \right. \\
	&	\left.   f(t)  \sqrt{\textbf{x}_{x}^{'2}(t) +\textbf{x}_{y}^{'2}(t)  }\right)
	\end{split}
	\end{equation}
	The resulting $\overline{\textbf{y}}(t)$ is a so-called RE curve, a curve yielding \underline{r}ational \underline{e}nvelope. For the construction, see Fig.~\ref{fig:bizzarri-ybar}. The authors in \cite{Bizzarri2016} prove that the envelope is indeed rational. Moreover, if $\textbf{x}(t)$ is a PH curve, then $\overline{\textbf{y}}(t)$ is an MPH curve, and the offsets of the envelope are rational as well. 
	
	\subsection{Skinning of circles}\label{subsection:skinning}
	
\begin{figure}
		\centering
		\includegraphics[width=.5\linewidth,, trim={5cm 1cm 29cm 2cm},clip]{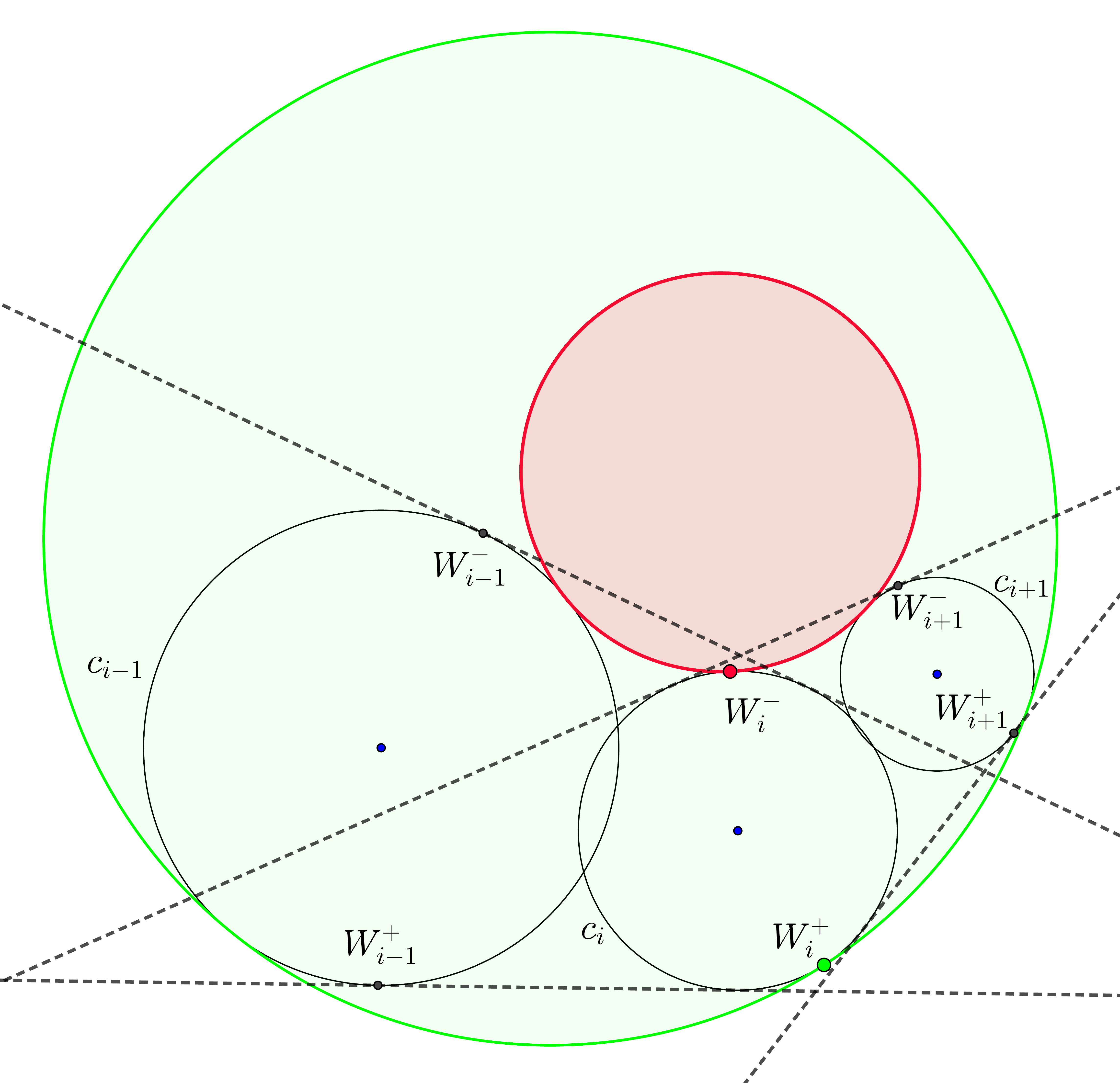}
		\caption{Determining touching points with the method of \cite{Kunkli2010} for a sequence of circles. For each circle  $c_i$, the touching points are determined by using the solutions to the Apollonius problem. In the case of the first and last circle, the outer common tangents are used.}
		\label{fig:apollonius}
	\end{figure}

	As mentioned in Section \ref{section:introduction}, skinning is a technique used in Computer Aided Geometric Design for modeling. Now let us introduce the skinning method of \cite{Kunkli2010}. Given an ordered set of circles, skinning is the construction of two G\textsuperscript{1} continuous curves touching each of the given circles at one point separately. One of the most important steps is the localization of the touching points (see Fig.~\ref{fig:apollonius}). The touching points for the first and last circles are determined by the two common outer tangents of the circles, and for the other circles, circle triplets are considered. For each triplet, we find two special solutions to the problem of Apollonius, by which we can define the touching points for the middle circle. Once we determine the points of contact, then they are separated into two groups for the ``left'' and for the ``right'' skin. Finally, Hermite interpolation curves are constructed between every two neighboring circles, for which the tangent lengths are computed with the help of the radical lines of the circle pairs. The resulting $G^1$ continuous splines are called the skins of the circle set.
	
\begin{figure}[t]
		\centering
		\includegraphics[width=.5\linewidth,, trim={2cm 9cm 11cm 5cm},clip]{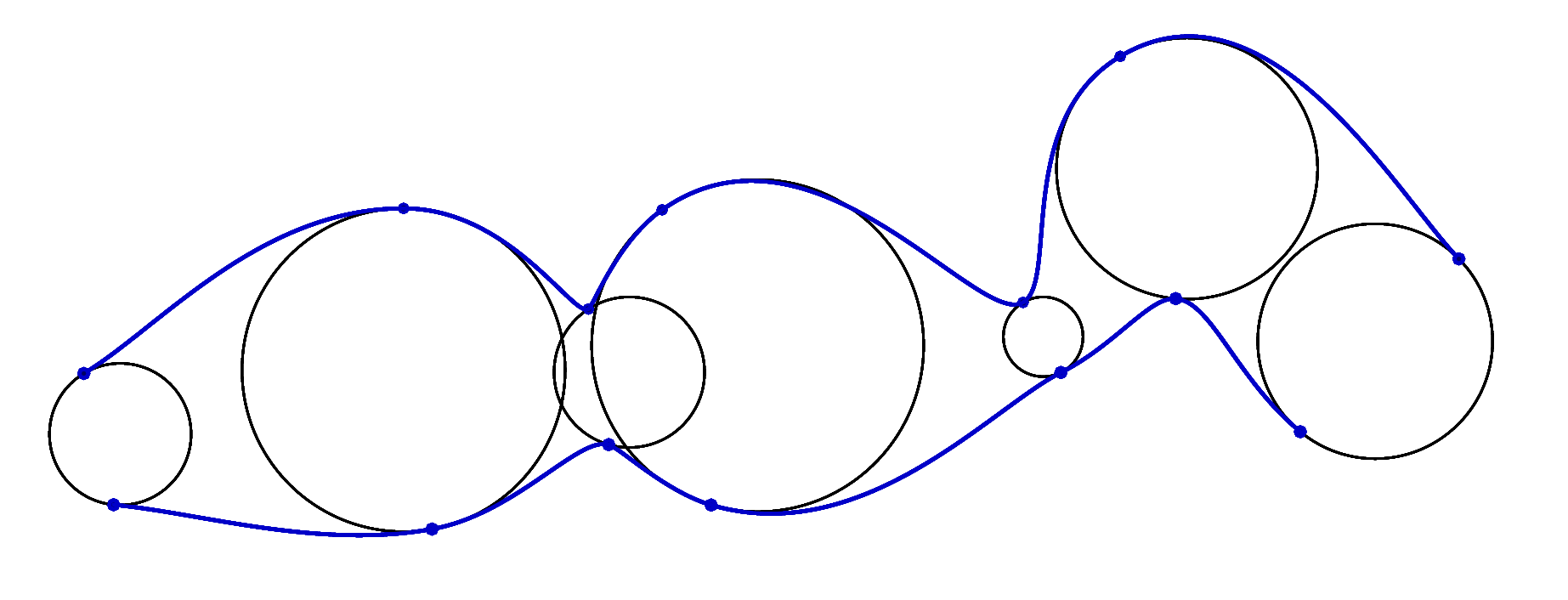}
		\caption{Skinning circles with the method given by \cite{Kunkli2010}.}
		\label{fig:skinnig-example}
	\end{figure}

	The algorithm generally provides good results for various input sets, and the skins respond dynamically when the user modifies the positions or the radii of the circles during modeling. Since the touching points are derived from the solution to the problem of Apollonius, it also guarantees that they never lie inside any of the neighboring circles. For an example, see Fig.~\ref{fig:skinnig-example}.
	
	\section{Applying MPH/RE curves for skinning}\label{section:oursolution}
	
	In this section, we propose a new application area for MPH/RE curves: skinning discrete sets of circles. As the problem setting of skinning is to find the bounding curves for an admissible set of predefined circles, we can define the appropriate input set of circles as in \cite{Kruppa2019}.
	
\begin{figure}
		\centering
		\includegraphics[width=0.5\linewidth, trim={2cm 6cm 10cm 4.5cm},clip]{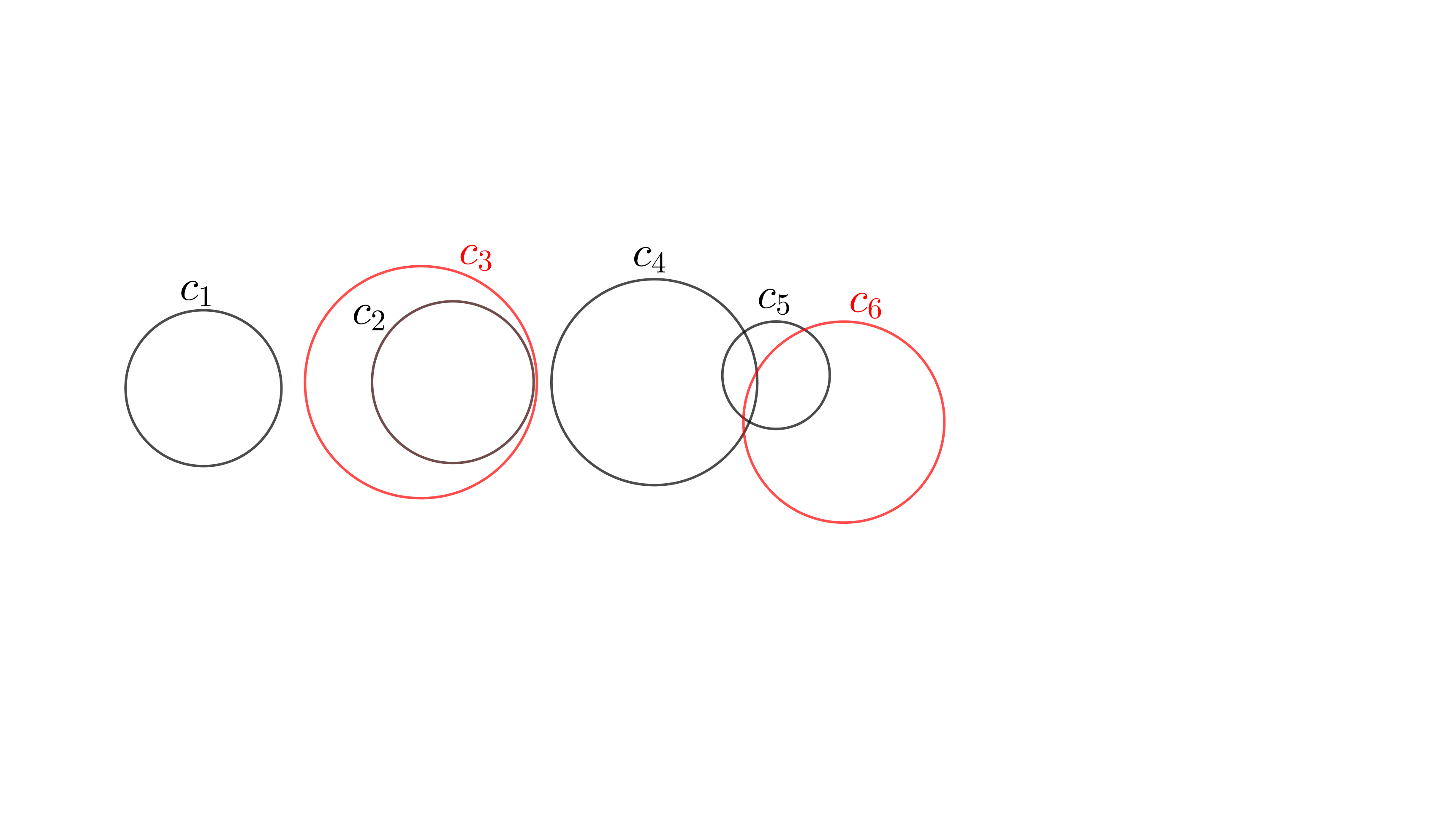}
		
		\includegraphics[width=0.45\linewidth, trim={2.3cm 5.7cm 14cm 2.5cm},clip]{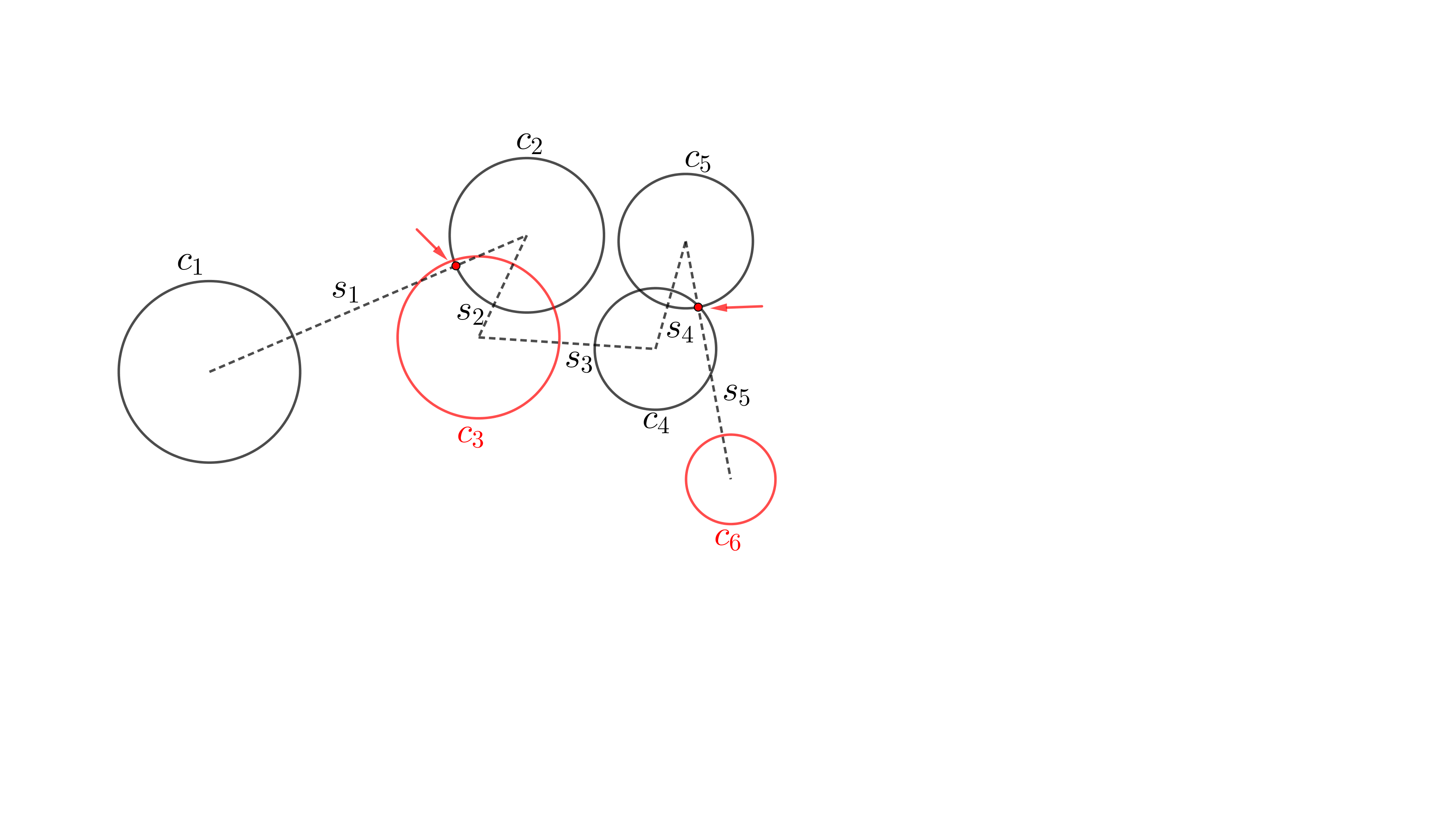}
		\caption{These example input sets are not admissible configurations, as the red circles are in incorrect positions. The positions of the red circles has to be changed according to Definition~\ref{def:admissible}.}
		\label{fig:admissible}
	\end{figure}

	\begin{definition}\label{def:admissible}
		Given an ordered set of circles $ \mathcal{C} = \left\lbrace c_1, c_2, \ldots, c_n \right\rbrace, n \in \mathbb{N}$, and the corresponding disks $\mathcal{D} = \left\lbrace d_1, d_2, \ldots, d_n \right\rbrace$, $\mathcal{C}$ is an admissible configuration for skinning, if it satisfies the following conditions (see Fig.~\ref{fig:admissible}):
		
		\begin{itemize}
			\item $d_{i} \not\subset \bigcup\limits_{j=1, j \neq i}^{n} d_{j},\\  i \in \left\lbrace 1, 2, \ldots, n \right\rbrace$
			\item $d_{i} \cap d_{j} = \emptyset$,
			\\
			$ i, j \in \left\lbrace 1, 2, \ldots, n \right\rbrace,\\ j \notin \left\lbrace i-2, i-1, i, i+1, i+2 \right\rbrace $
			\item $d_{i-1} \cap d_{i+1} \neq \emptyset \implies d_{i-1} \cap d_{i+1} \subset d_i \\ i \in \left\lbrace 2, 3, \ldots, n-1 \right\rbrace$ 
			\item $s_{i} \cap c_{i+2} \notin c_{i+1},\quad i \in \left\lbrace 1, 2, \ldots, n-2 \right\rbrace$ 
			\item $s_{i} \cap c_{i} \notin c_{i-1},\quad i \in \left\lbrace 2, 3, \ldots, n \right\rbrace$.
		\end{itemize}
	\end{definition}
	
	To attain a rational envelope for an admissible sequence of circles $\mathcal{C}$---by applying the RE interpolation method of \cite{Bizzarri2016}---for each circle pair we need to construct the input data for the interpolation. 
	
	Using cyclographic mapping, we can map the circles to points in $\mathbb{R}^{2,1}$, so that we obtain a sequence of points $\mathcal{P}=\{P_1, P_2, \dots, P_n\}$, where for each circle $c_i$---with center $O_i (O_{i_x}, O_{i_y})$ and radius $r_i$---the corresponding spatial point is $P_i = ( O_{i_x}, O_{i_y}, r_i )$. However, since we do not have any additional information about the circles, there are no initial tangent vectors provided. 
	
\begin{figure}
		\centering
		\includegraphics[width=.8\linewidth,, trim={10cm 13.5cm 3cm 9.5cm},clip]{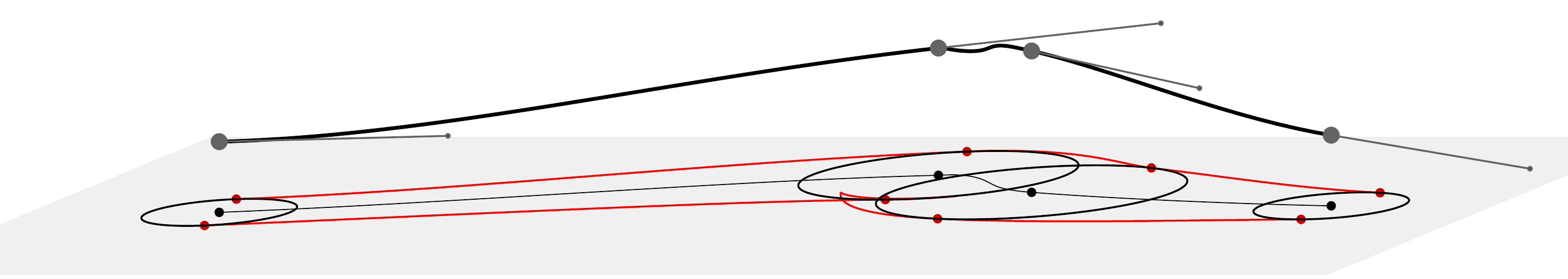}
		\includegraphics[width=.8\linewidth,, trim={10cm 15cm 7cm 4cm},clip]{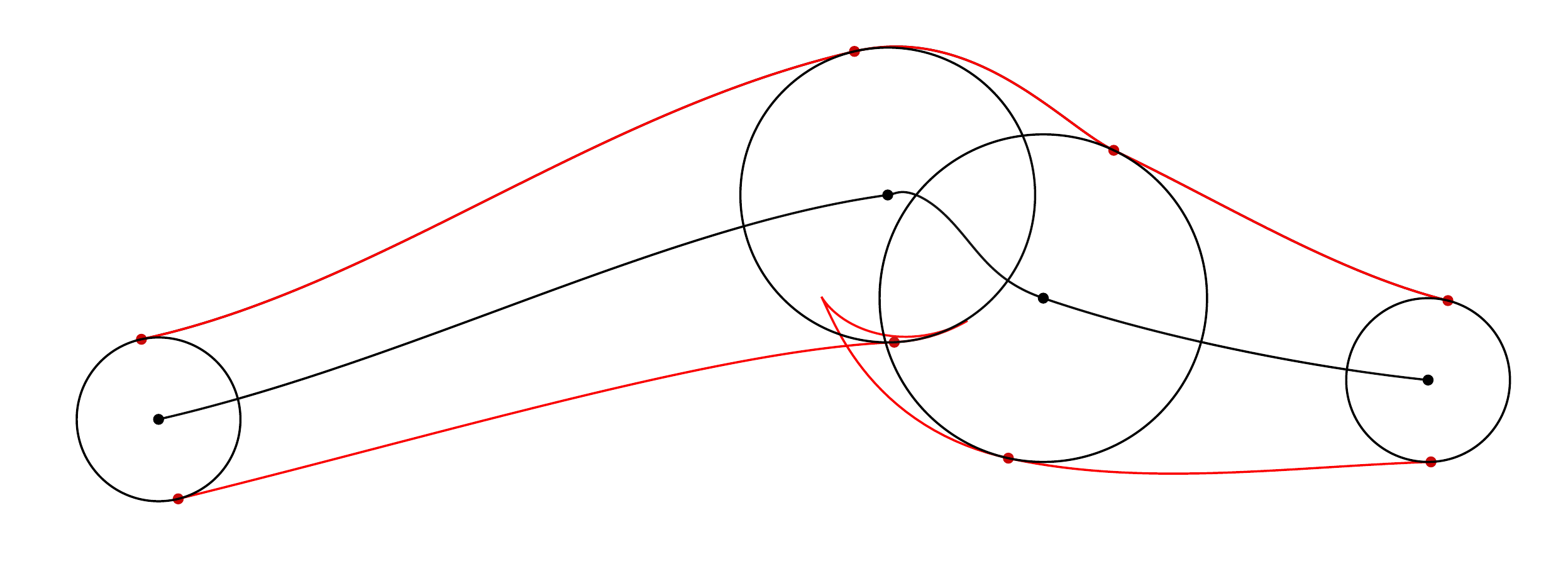}
		\caption{By using seemingly appropriate tangent vectors that are chosen freely, we can observe that the touching points can get inside the circles.}
		\label{fig:tp-bad}
	\end{figure}

	We may define the vectors using e.g., the Catmull-Rom spline interpolation, so that each $\textbf{t}_i$ tangent vector is defined as $\textbf{t}_i =\lambda  \cdot (P_{i+1}-P_{i-1})$, ($i \in [2,n-1]$, $\lambda \in \mathbb{R}^{+}$). For the first and last ones, the vectors are defined as $\textbf{t}_1 = \lambda \cdot (P_{2}-P_{1})$ and $\textbf{t}_n = \lambda \cdot (P_{n}-P_{n-1})$. Then, by using Formula \ref{eq:env_q} we calculate the positions of the touching points, and by constructing the $\overline{\textbf{y}}$ RE curve (see Formula~\ref{eq:re_curve}), we obtain the envelope. Generally this provides adequate output, however, in the case of intersecting circles, the resulting touching point can get inside the neighboring circle, and thus, the envelope can cut into the circles. As an example, refer to Fig.~\ref{fig:tp-bad}.
	
	To resolve this problem, we can approach from an inverse aspect. As we could see in Section \ref{section:previouswork}, the method of \cite{Kunkli2010}  localizes the touching points by solving the problem of Apollonius, which guarantees that the points never lie inside any of the neighboring circles. Our idea is to localize the touching points at first, and then reconstruct that tangent vector with which we would obtain exactly these points.
	
	\subsection{Reconstructing tangent vectors}
	
	Regarding every circle $c_i$ in $\mathcal{C}$, let us fix two points on the circle which we want to use later as the touching points of the envelope. Let us mark these points as $W^+_i$ and $W^-_i$. The aim is to determine that tangent vector for which we get $W^\pm_i = Q^\pm_i$ using the Formula~\ref{eq:env_q}.

	It is known that the envelope formula describes a geometric construction (\cite{Peternell1998, Peternell2008}), and \cite{Kunkli2009} also presented different ways to obtain touching points for skinning, one of which was using this geometrical approach to create the corresponding touching points. We used this as an inspiration to reconstruct the tangent vector $\textbf{t}_i$ from the touching points.
	
	Let $W^-_i$ and $W^+_i$ note the left and right touching points on $c_i$ as shown in Fig.~\ref{fig:apollonius}, and we aim to construct $S_i$, the endpoint of the desired tangent vector. Let us define line $e_i = \overleftrightarrow{W^+_i W^-_i}$ and line $f_i$ so that $\overset{\triangledown}{P_i}\in f_i$ and $f_i \perp e_i$. Let $I_i$ denote the intersection point of $e_i$ and $f_i$. $S_i$ now can be defined as the inverse point of $I_i$ with respect to the circle $c_i$. 
	\begin{align}\label{eq:s}
	S_{i_x} &= P_{i_x} + r_i^2  \frac{W_{i_x}^+  - W_{i_x}^-   }{\gamma_i}
	\\
	S_{i_y} &= P_{i_y} - r_i^2  \frac{W_{i_y}^+  - W_{i_y}^-   }{\gamma_i}
	\\
	\gamma_i &= P_{i_x}   \left(  W_{i_y}^- - W_{i_y}^+\right)   + P_{i_y}   \left(  W_{i_x}^- - W_{i_x}^+\right)  + \notag\\
	&\phantom{{}={}}  W_{i_y}^+  W_{i_x}^-   - W_{i_x}^+  W_{i_y}^-  
	\end{align}
	
	Once $S_i$ is obtained, the direction of the tangent vector $\textbf{t}_i$ still has to be defined. Let us construct the vector $\textbf{v}^+_i$ by rotating the vector $W^+_i-\overset{\triangledown}{P_i}$  counterclockwise by $\frac{\pi}{2}$, and let $\textbf{u}_i$ be defined as $S_i-P_i$. Since we would like the tangent vector $\textbf{t}_i$ to point towards the next circle, we can choose between the two directions by measuring the angle between $\textbf{v}^+_i$ and $\overset{\triangledown}{\textbf{u}_i}$, and we can define $\textbf{t}_i$ as follows:
	
	\begin{equation}\label{tangent_direction}
	\textbf{ t}_i=
	\begin{cases}
	\frac{S_i-P_i}{||S_i-P_i||}, & \text{if}\  \textbf{v}^+_i \cdot \overset{\triangledown}{\textbf{u}_i} > 0  \\
	\frac{P_i-S_i}{||S_i-P_i||}, & \text{otherwise}
	\end{cases}
	\end{equation}

\begin{figure}[hb!]
		\centering
		\includegraphics[width=.8\linewidth,, trim={0 0 0 0},clip]{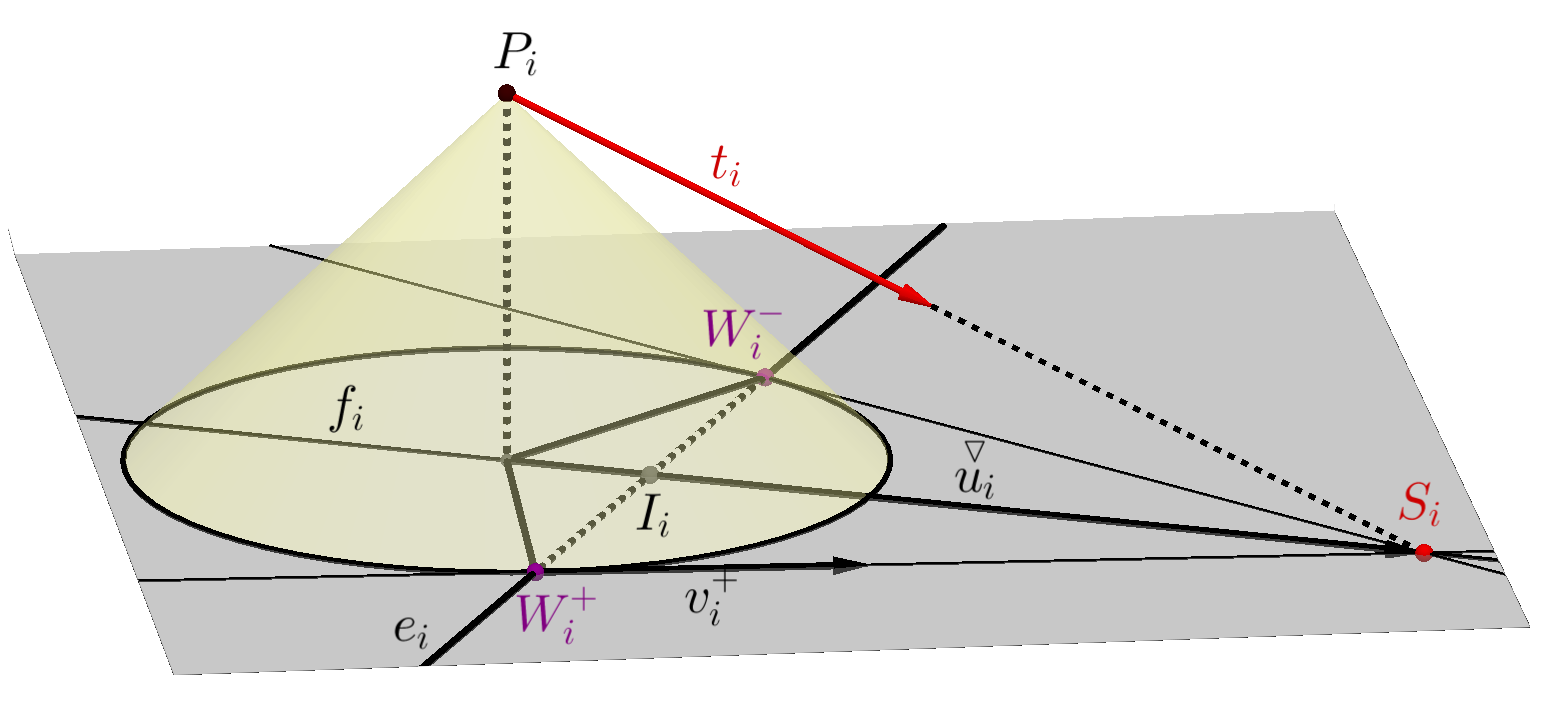}
		\caption{Given the touching points $W^+_i$ and $W^-_i$, we can reconstruct the tangent vector $\textbf{t}_i$. The construction is based on the geometric properties of the envelope of a family of circles. }
		\label{fig:tp-geometric}
	\end{figure}

	To see the geometric construction of $\textbf{t}_i$, we refer the reader to Fig.~\ref{fig:tp-geometric}. Since our construction is exactly the inverse of the standard geometric construction of the envelope points, it is trivial that once we define $\textbf{t}_i$ this way, the obtained $Q^+_i$ and $Q^-_i$ points using Formula~\ref{eq:env_q} are the same as $W^+_i$ and $W^-_i$. Also, it is important to note that choosing the tangent vector this way assures that $\textbf{t}_i$ is always space-like.

	\subsection{Constructing the envelope curves}
	
	We have seen that the RE method defined by \cite{Bizzarri2016} uses the $\alpha_i$ values in Formula~\ref{eq:env_v} as free parameters which essentially affect the shape of the resulting envelope, as $\overline{\textbf{y}}(t)$ is constructed from the planar curve $\textbf{x}(t)$. Since now we use the touching points of \cite{Kunkli2010}, it is natural to choose the tangent length the authors suggested. Let $l_i$ denote the radical line of $c_i$ and $c_{i+1}$. To assure an aesthetically more satisfying result, for each neighboring circle pair, the starting and ending tangent lengths will be different. Thus for the circle pair $c_i$ and $c_{i+1}$, the corresponding tangent vectors $\textbf{v}_i$ and $\textbf{v}_{i+1}$ for the touching point are defined as follows:
	\begin{equation}\label{eq:v}
	\textbf{v}_j = \alpha_j \ \frac{{(\overset{\triangledown}{P_i} - Q_i)}^\perp}{||{\overset{\triangledown}{P_i} - Q_i}||}
	\qquad j \in \{i,i+1\}
	\end{equation}
	\noindent where $Q_j = Q^+_j = W^+_j$. The $\alpha_j$ values can be determined using the perpendicular distance:
	$$\alpha_j = 2 \ \text{dist}(Q_j, l_i)$$
	To define the medial axis, we only need to follow the remaining steps of the algorithm in \cite{Bizzarri2016}. After having $\textbf{x}(t)$ defined as a rational planar curve interpolating $Q_i$, $Q_{i+1}, \textbf{v}_i,$ and $\textbf{v}_{i+1}$ (with e.g., Hermite arc), we construct the polynomial function $f$ with the boundary conditions (Formula~\ref{eq:fcondition}), and we obtain the final RE curve in the form of Formula~\ref{eq:re_curve}.
	The resulting $\overline{\textbf{y}}$ is an RE curve, for which by applying Formula~\ref{eq:envelope}, the resulting envelope curves are rational.

\begin{figure}[h!]
		\centering
		\includegraphics[width=.8\linewidth,, trim={10cm 13.5cm 3cm 9.5cm},clip]{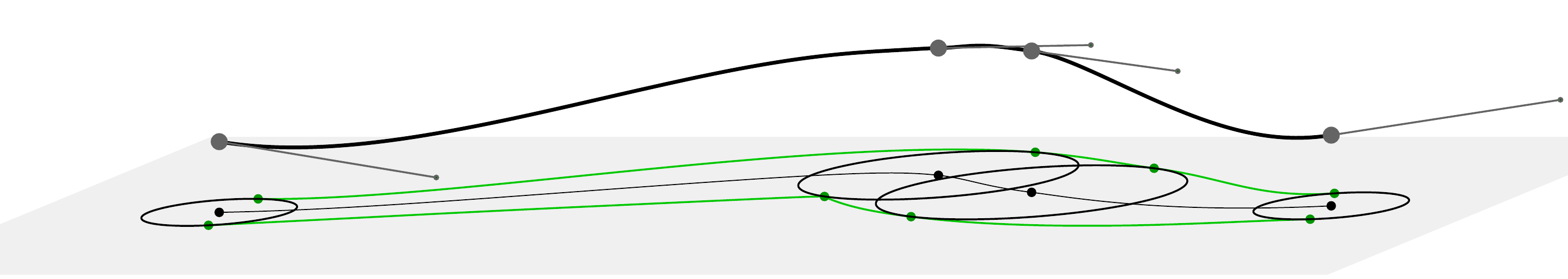}
		\includegraphics[width=.8\linewidth,, trim={10cm 15cm 7cm 4cm},clip]{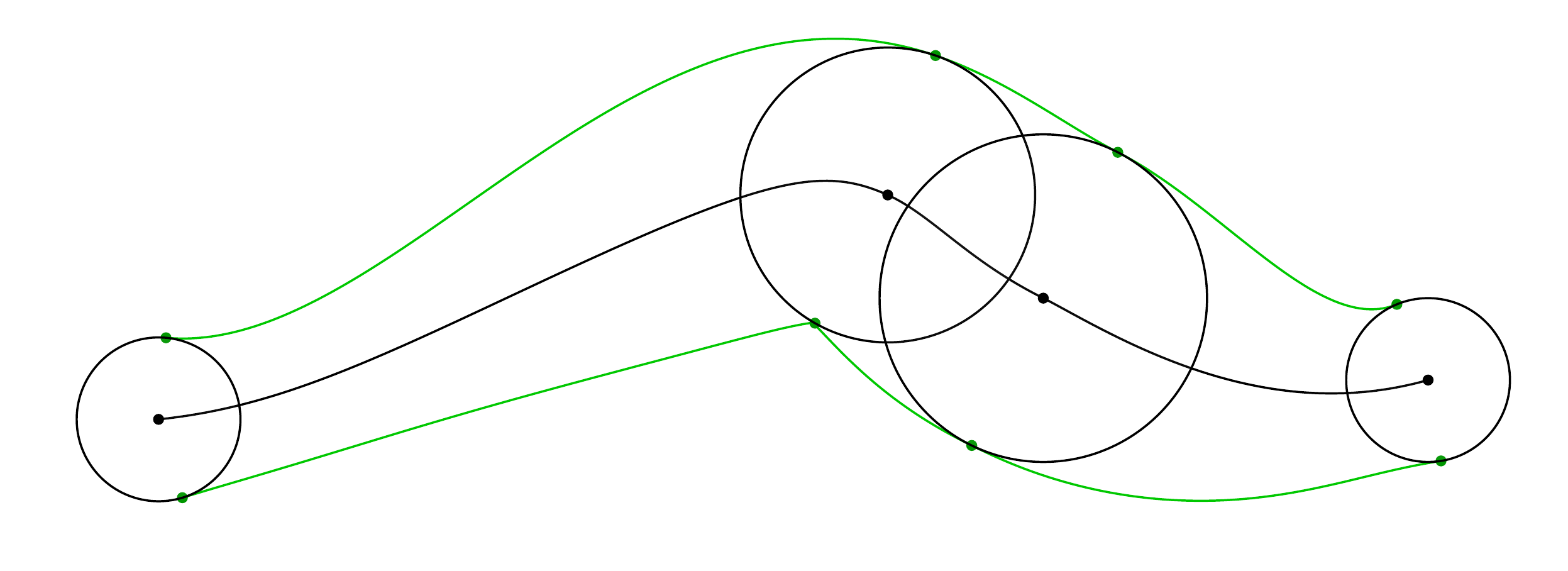}
		\caption{For the same input as in Fig.~\ref{fig:tp-bad}, we can observe that the touching point now lie outside the neighboring circle. If we reconstruct the tangent vectors from the previously chosen touching points, we are able to use the RE curve for skinning. }
		\label{fig:tp-good}
	\end{figure}

\begin{figure}[ht]
		\centering
		\includegraphics[width=.8\linewidth, trim={9cm 12cm 6cm 9cm},clip]{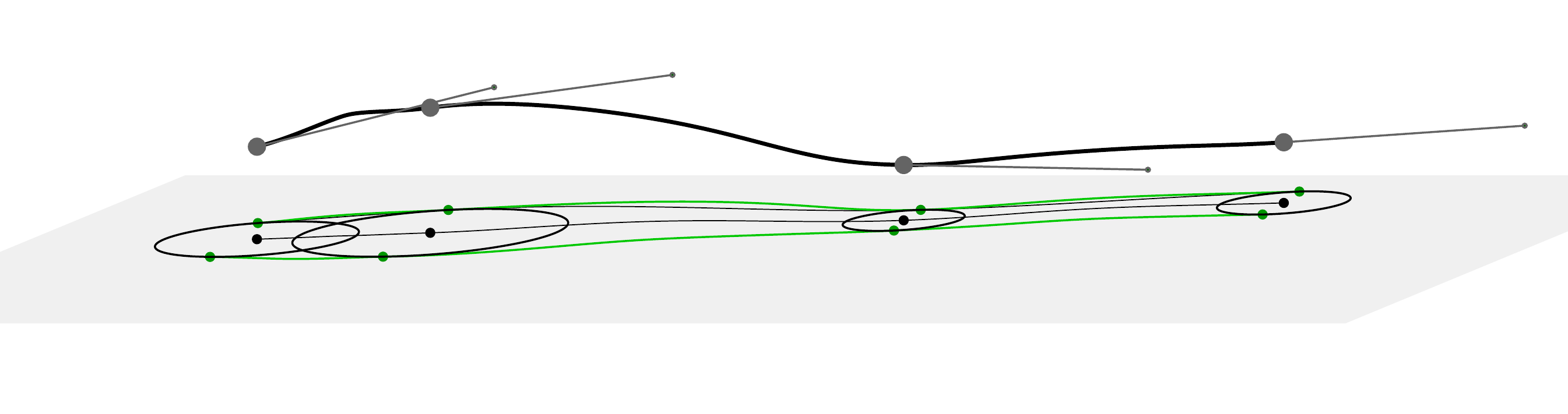}
		\includegraphics[width=.75\linewidth,, trim={9cm 14cm 13cm 1cm},clip]{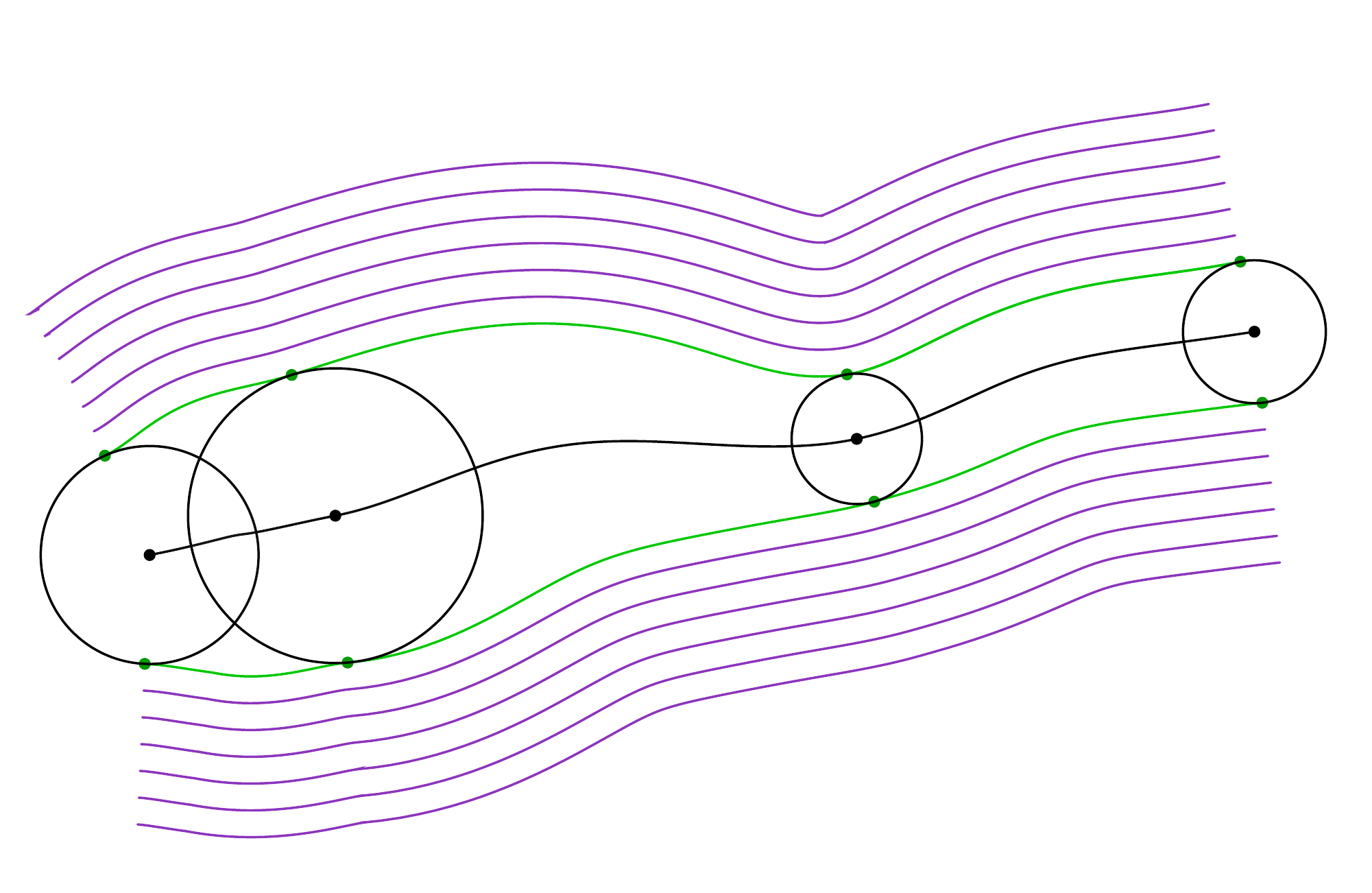}
		\caption{MPH curve used for skinning with our inverse calculation approach. Not only the envelope curves are rational, but the offsets too. }
		\label{fig:offset}
	\end{figure}

	Fig.~\ref{fig:tp-good} shows our results using the inverted calculation to define the Hermite input data, so that the resulting touching points of the envelope are always adequate for the purpose of skinning. \cite{Bizzarri2016} proved that if we construct $\textbf{x}(t)$ as a PH curve, not only the envelope is rational, but its offsets as well. Fig.~\ref{fig:offset} shows such a scenario: the resulting offsets are rational too.

	\section{Conclusions}\label{section:conclusion}

	RE curves were introduced by \cite{Bizzarri2016}, and they presented an interpolation method which can be used to create rational surface blends for canal surfaces. In this paper, we propose a new application for RE curves: skinning a discrete set of circles. If we aim to apply the algorithm to obtain the envelope bounding a discrete set of circles, we face a serious obstacle: even though the circles can be regarded as spatial points using cyclographic mapping, there is no kind of information for the tangent vectors that should be used for the interpolation. Obviously, the shape of the envelope fundamentally depends on the chosen tangent vectors. Even by choosing a seemingly suitable method (e.g., the Catmull-Rom interpolation) to define these vectors, we can see that the resulting touching points and the shape of the envelope is sometimes problematic. When two circles intersect, the touching points of the circle can lie inside other circles. To resolve this problem, we offer an inverted approach: we predefine the positions of the touching points (with the method of \cite{Kunkli2010}), and then we calculate which is the appropriate tangent vector that would provide these points as touching points. We also fix the free parameters in the method of \cite{Bizzarri2016} by giving an exact calculation, which also greatly affects the shape of the envelope. This way our method does not need a numerical optimization afterwards to generate a good output. 
	
	To sum up, we offer a new application for RE curves: skinning a discrete set of circles which provides a rational envelope (and can be constructed so that the offsets are rational as well). We give an exact solution, thus, we guarantee that there exists only one envelope for a user-defined circle sequence. With our method, we solve the problem when the touching points of a circle could get inside the neighboring circles. As a future work, we plan to extend our algorithm to 3D to provide another application for RE curves: skinning a discrete set of spheres. 
	
	\bibliographystyle{unsrt}  

\begin{thebibliography}{10}
		
		\bibitem{Bizzarri2016}
		Michal Bizzarri, Miroslav L{\'{a}}vi{\v{c}}ka, and Ji{\v r}{\'{i}} Kosinka.
		\newblock {Medial axis transforms yielding rational envelopes}.
		\newblock {\em Computer Aided Geometric Design}, 46:92--102, 2016.
		
		\bibitem{Choi1997}
		Hyeong~In Choi, Sung~Woo Choi, and Hwan~Pyo Moon.
		\newblock {Mathematical theory of medial axis transform}.
		\newblock {\em Pacific Journal of Mathematics}, 181(1):57--88, 1997.
		
		\bibitem{Choi1999}
		Hyeong~In Choi, Chang~Yong Han, Hwan~Pyo Moon, Kyeong~Hah Roh, and Nam~Sook
		Wee.
		\newblock {Medial axis transform and offset curves by Minkowski Pythagorean
			hodograph curves}.
		\newblock {\em Computer Aided Design}, 31(1):59--72, 1999.
		
		\bibitem{Moon1999}
		Hwan~Pyo Moon.
		\newblock {Minkowski Pythagorean hodographs}.
		\newblock {\em Computer Aided Geometric Design}, 16(8):739--753, 1999.
		
		\bibitem{Kim2003}
		Gwang-Il Kim and Min-Ho Ahn.
		\newblock {$C^1$ Hermite interpolation using MPH quartic}.
		\newblock {\em Computer Aided Geometric Design}, 20(7):469--492, 2003.
		
		\bibitem{Kosinka2006}
		Ji{\v r}{\'{i}} Kosinka and Bert J{\"{u}}ttler.
		\newblock {Hermite interpolation by Minkowski Pythagorean hodograph cubics}.
		\newblock {\em Computer Aided Geometric Design}, 23(5):401--418, 2006.
		
		\bibitem{Kosinka2009}
		Ji{\v r}{\'{i}} Kosinka and Bert J{\"{u}}ttler.
		\newblock {$C^1$ Hermite interpolation by Pythagorean hodograph quintics in
			Minkowski space}.
		\newblock {\em Advances in Computational Mathematics}, 30(2):123--140, 2009.
		
		\bibitem{Kosinka2010}
		Ji{\v r}{\'{i}} Kosinka and Zbyn{\v{e}}k {\v{S}}{\'{i}}r.
		\newblock {$C^2$ Hermite interpolation by Minkowski Pythagorean hodograph
			curves and medial axis transform approximation}.
		\newblock {\em Computer Aided Geometric Design}, 27(8):631--643, 2010.
		
		\bibitem{Kosinka2011}
		Ji{\v r}{\'{i}} Kosinka and Miroslav L{\'{a}}vi{\v{c}}ka.
		\newblock {A unified Pythagorean hodograph approach to the medial axis
			transform and offset approximation}.
		\newblock {\em Journal of Computational and Applied Mathematics},
		235(12):3413--3424, 2011.
		
		\bibitem{Bizzarri2019}
		Michal Bizzarri, Miroslav L{\'{a}}vi{\v{c}}ka, and Jan Vr{\v{s}}ek.
		\newblock {Linear computational approach to interpolations with polynomial
			Minkowski Pythagorean hodograph curves}.
		\newblock {\em Journal of Computational and Applied Mathematics}, 361:283--294,
		2019.
		
		\bibitem{Kunkli2010}
		Roland Kunkli and Mikl{\'{o}}s Hoffmann.
		\newblock {Skinning of circles and spheres}.
		\newblock {\em Computer Aided Geometric Design}, 27(8):611--621, 2010.
		
		\bibitem{Bana2014}
		Korn{\'{e}}l Bana, Kinga Kruppa, Roland Kunkli, and Mikl{\'{o}}s Hoffmann.
		\newblock {KSpheres -- an efficient algorithm for joining skinning surfaces}.
		\newblock {\em Computer Aided Geometric Design}, 31(7-8):499--509, 2014.
		
		\bibitem{Bastl2015}
		Bohum{\'{i}}r Bastl, Ji{\v r}{\'{i}} Kosinka, and Miroslav L{\'{a}}vi{\v{c}}ka.
		\newblock {Simple and branched skins of systems of circles and convex shapes}.
		\newblock {\em Graphical Models}, 78:1--9, 2015.
		
		\bibitem{Kruppa2019}
		Kinga Kruppa, Roland Kunkli, and Mikl{\'{o}}s Hoffmann.
		\newblock {An improved skinning algorithm for circles and spheres providing
			smooth transitions}.
		\newblock {\em Graphical Models}, 101:27--37, 2019.
		
		\bibitem{Peternell1998}
		Martin Peternell and Helmut Pottmann.
		\newblock {Applications of Laguerre geometry in CAGD}.
		\newblock {\em Computer Aided Geometric Design}, 15(2):165--186, 1998.
		
		\bibitem{Peternell2008}
		Martin Peternell, Boris Odehnal, and Maria~Lucia Sampoli.
		\newblock {On quadratic two-parameter families of spheres and their envelopes}.
		\newblock {\em Computer Aided Geometric Design}, 25(4-5):342--355, 2008.
		
		\bibitem{Kunkli2009}
		Roland Kunkli.
		\newblock {Localization of touching points for interpolation of discrete
			circles}.
		\newblock {\em Annales Mathematicae et Informaticae}, 36(1):103--110, 2009.
		
	\end{thebibliography}

\end{document}